\documentclass[prd,preprint,superscriptaddress,showpacs,byrevtex]{revtex4}

\usepackage{epsfig}

\newcommand{\be}{\begin{equation}}
\newcommand{\ee}{\end{equation}}
\newcommand{\bea}{\begin{eqnarray}}
\newcommand{\eea}{\end{eqnarray}}

\begin{document}

\begin{flushright}
YITP-SB-05-32
\end{flushright}

\title{$J/\Psi$ and $\Psi^\prime$ Polarizations in Polarized Proton-Proton\\
Collisions at the RHIC}

\author{Gouranga C. Nayak} \email{nayak@insti.physics.sunysb.edu} 
\affiliation{
C. N. Yang Institute for Theoretical Physics, Stony Brook University, 
SUNY, Stony Brook, NY 11794-3840, USA }
\author{J. Smith} \email{smith@insti.physics.sunysb.edu} 
\affiliation{
C. N. Yang Institute for Theoretical Physics, Stony Brook University, 
SUNY, Stony Brook, NY 11794-3840, USA\\
and\\
NIKHEF, Postbus 41882, 1009 DB, Amsterdam, The Netherlands }

\date{\today}
\begin{abstract} 

We study inclusive heavy quarkonium production with definite 
polarizations in polarized proton-proton collisions using the 
non-relativistic QCD color-octet mechanism.  We present results for 
rapidity distributions of cross sections and spin asymmetries for the 
production of $J/\psi $ and $\psi^\prime$ with specific
polarizations in polarized p-p collisions at $\sqrt s$ = 200 GeV and 500 GeV 
at the RHIC within the PHENIX detector acceptance range.

\end{abstract} 

\maketitle
\newpage
\section{Introduction}

The relativistic heavy ion collider (RHIC) at BNL is a unique facility
which collides two heavy ions at $\sqrt s$ = 200 GeV 
to study the production of a quark-gluon plasma \cite{qgp}
and two polarized protons to study the parton spin structure of 
the proton \cite{spin}.
Measurements of heavy probes such as $J/\psi$ production
and Drell-Yan production are useful tools to study the quark-gluon plasma 
in heavy ion collisions and to extract the polarized gluon distribution 
functions inside the proton in polarized p-p collisions 
\cite{phenix,QWGBrambilla}.
Understanding the correct production mechanisms for these processes 
is important.  Heavy quarkonium production can be
described via the non-relativistic QCD (NRQCD) color-octet mechanism 
\cite{bodwin,nayak}. 
                                                                                
In NRQCD the energy eigenstates of heavy quarkonium bound states $|H>$
are labelled by the quantum numbers $J^{PC}$, with an additional
superscript to give the color; (1) for singlet and (8) for octet.
When the Fock states are analysed then the dominant component
in S-wave orthoquarkonium is the pure
quark-antiquark state
$|Q\bar Q[^3S^{(1)}_1]>$. A state with dynamical gluons, such as
$|Q\bar Q[^3P^{(8)}_J]g>$ does contribute but only with a probability
of order $v^2$,
where $v$ is the typical velocity of the non-relativistic heavy quark
(and antiquark). The other states, such as
$|Q\bar Q[^3S^{(1,8)}_1]gg>$, $|Q\bar Q[^1S^{(8)}_0]g>$ and
$|Q\bar Q[^3D^{(1,8)}_J]gg>$
contribute to the probability in even higher orders in $v$ (see later).
Correspondingly the dominant states in P-wave orthoquarkonia are the states
$|Q\bar Q[^3P^{(1)}_J]>$,
and the states with dynamical gluons such as $|Q\bar Q[^3S^{(8)}_1]g>$
contribute with a probability of order $v^2$.
After a $Q\bar Q$ is formed in a color octet state it may emit a soft gluon
to transform into the singlet state
$|Q\bar Q[^3P^{(1)}_J]>$ and then become a $J/\psi$ state by photon decay.
The $Q\bar Q$ pair in a color octet state can also emit two long
wavelength gluons and then become a $J/\psi$ state.
All these low energy interactions are negligible and the
non-perturbative matrix elements, labelled by the above quantum numbers,
can be fitted from experiments or
can be determined from lattice field theory calculations.

Using the NRQCD color octet mechanism heavy quarkonia production rates 
have been calculated for the p-$\bar {\rm p}$ Tevatron collider
\cite{CDFoctet,CL}, for the e-p HERA collider \cite{HERAoctet}, for the
e${}^+$-e${}^-$ LEP collider \cite{LEPoctet} 
and for fixed target experiments \cite{fixedtargetoctet}. 
Also the recent PHENIX data for $J/\psi$ production  
in unpolarized p-p collisions can be
explained by this same mechanism \cite{cooper}. 

The RHIC offers a wide variety of measurements 
with respect to $J/\psi$ production.
They involve $J/\psi$ production (with and without definite
polarizations), in unpolarized p-p, d-Au, Cu-Cu and Au-Au collisions
and in polarized p-p collisions.
Since the maximum transverse momentum of heavy quarkonium that can be measured
at the RHIC is around 10 GeV/c, the 
parton fragmentation contribution to heavy quarkonium production
will be very small and we will neglect it in our study. 
The main contributions to heavy quarkonium 
production at the RHIC are the parton fusion processes \cite{cooper,pol}.

The inclusive heavy quarkonium production cross section
(summed over quarkonium polarization states) in unpolarized and 
polarized partonic collisions were calculated 
in \cite{CL,fm} and \cite{pol,gm} respectively. 
Heavy quarkonium production cross sections with definite polarizations in 
unpolarized partonic collisions were calculated 
in \cite{braaten,CDFpolarization}.
In this paper we will study the inclusive rapidity distributions
for heavy quarkonium production with specific polarizations in polarized p-p
collisions. We will evaluate the partonic level cross sections 
for the processes $q \bar q, ~gg ~\rightarrow ~
J/\psi (\lambda) (\psi^\prime(\lambda))$ in polarized p-p collisions where
$\lambda$ is the helicity (polarization) of the heavy quarkonium state. 
This study should be regarded as a preliminary analysis of the leading order
(LO) contributions to charmonium production in specific polarization states. 
The factorization ansatz, which separates the short-distance coefficient 
functions, calculable in perturbative QCD, from the long distance matrix 
elements, which are fitted to data, is only proved if the charmonium state
is produced at a large transverse momentum \cite{nayak}. 
Our LO analysis considers only 
charmonium production in the forward direction at a finite rapidity.
It will be followed later by a next-to-leading (NLO) calculation where 
additional quark and/or gluon radiation is produced in the final state so that
the charmonium state does have a finite transverse momentum.     
The reason we need these results is that the PHENIX collaboration at the 
RHIC will measure $J/\psi$ and $\psi^\prime$ production with definite 
polarizations in polarized p-p collisions at $\sqrt s$ = 200 GeV and 500 GeV
\cite{phenix}.  Since polarized heavy quarkonium production at the 
Tevatron energy scale \cite{expt} is not explained by the NRQCD color-octet 
mechanism \cite{CDFpolarization} it will
be useful to compare our results for $J/\psi$ and $\psi^\prime$ polarizations 
with the future data at the RHIC. The study of polarized heavy quarkonium 
production in polarized p-p collisions at the RHIC is also unique in the 
sense that it probes the spin transfer processes in perturbative QCD (pQCD).
Note that decays from higher quarkonium states to the $J/\psi$ are ignored.
We only consider direct production.

The spin projection method is used to evaluate the inclusive cross section for 
heavy quarkonium production (summed over polarization states) 
in parton fusion processes \cite{CL}. However, the cross section for 
heavy quarkonium production with a specific polarization in the final state 
can involve additional matrix elements that do not contribute when the 
polarization is summed. This includes interference terms between 
partonic processes that produce heavy quark-antiquark 
pairs with different total angular momenta. 
Such interference terms cancel upon summing over polarizations.
These interference terms can be calculated by using the helicity decomposition 
method \cite{braaten}. Hence we will use the helicity decomposition
method to calculate the square of the matrix elements 
for heavy quarkonium production
with definite helicity in polarized partonic collisions. Using
these results we will compute the rapidity distributions of the cross sections
and spin asymmetries of heavy quarkonium production with definite helicity 
states in polarized p-p collisions at RHIC at $\sqrt s$ = 200 GeV and 500 GeV 
within the PHENIX detector acceptance ranges.

The paper is organized as follows. In section II we derive the partonic 
level cross sections for heavy quarkonium production with definite
polarizations in polarized q-$\bar{\rm q}$ and g-g parton fusion processes 
using the helicity decomposition method within the NRQCD color-octet mechanism.
In section III
we present the results for the differential rapidity distributions 
and spin asymmetries for the $J/\psi(\lambda)$ and $\psi^\prime(\lambda)$
in the PHENIX detector acceptance range in polarized p-p collisions 
at $\sqrt s$ = 200 GeV and 500 GeV. 
We then discuss these results and give our conclusions.

\section{Inclusive Heavy Quarkonium Production with Definite\\
Helicities in Polarized Partonic Collisions}

In this section we will use the NRQCD color-octet mechanism
and derive the square of the matrix element 
for inclusive heavy quarkonium production with definite helicities
in polarized partonic fusion processes.
We will consider the (polarized) partonic fusion processes
$q \bar q~\rightarrow ~ H(\lambda)$ and $gg ~\rightarrow ~ H(\lambda)$ 
where $\lambda$ is the helicity of the produced heavy quarkonium state $H$. 
We will use the helicity decomposition method \cite{braaten}
within the NRQCD color-octet mechanism to calculate these processes where 
both initial and final state particles are polarized. 

\subsection{The $q\bar q$ fusion process}

The production of a heavy charmonium state with helicity $\lambda$,
where $\lambda= 0, \pm 1$ correspond to longitudinal and transverse
polarization states respectively, begins with the calculation of the production
of a heavy quark anti-quark pair.
The matrix element for the light quark-antiquark ($q\bar q$) fusion 
process $q(k_1) + \bar q(k_2) \rightarrow Q(p_1) + \bar Q(p_2)$ 
producing a heavy quark-antiquark ($Q\bar Q$) pair is given by
\bea
M_{q \bar q \rightarrow Q\bar Q}~=~\frac{g^2}{P^2}~ \bar{v}(k_2) \gamma_\mu T^a
u(k_1) \, \bar{u}(p_1) \gamma^\mu T^a v(p_2)\,,
\label{eq1}
\eea
where $P^\mu~=~p_1^\mu~+~p_2^\mu~=~k_1^\mu~+~k_2^\mu$ and 
$p_1^\mu=P^\mu/2+L^\mu_jq^j$ and $p_2^\mu=P^\mu/2-L^\mu_jq^j$.
Here $P^\mu$ is the CM momentum of the pair and $q^i$ is their
relative momentum in the CM frame. The latter vector does not have any 
time component so $i=1,2,3$ only. 
$L^\mu_j$ is the boost matrix defined in \cite{braaten} with both Lorentz and 
three vector indices.
In terms of non-relativistic heavy quark Pauli spinors ($\xi$ and $\eta$) 
we obtain (up to terms linear in $q$):
\bea
|M_{q \bar q \rightarrow Q\bar Q}|^2~=~\frac{g^4}{4m^2}~ 
{\eta^\prime}^{\dagger} \sigma^i T^a {\xi^\prime} \,L^\mu_i\,
\bar{u}(k_1) \gamma_\mu T^a
v(k_2) 
\,\bar{v}(k_2) \gamma_\nu T^b
u(k_1)\, 
L^\nu_j \,{\xi}^{\dagger} \sigma^j T^b {\eta}\,,
\label{eq4}
\eea
where $m$ is the mass of the heavy quark. For massless incoming quarks and 
antiquarks we have
\bea
&& u(k_1) \bar{u}(k_1) ~=~ \frac{1}{2}~(1+ h_1\gamma_5)~ 
\gamma_\mu k_1^\mu  
\nonumber \\
&&~v(k_2) \bar{v}(k_2) ~=~ \frac{1}{2}~(1- h_2\gamma_5)~ 
\gamma_\mu k_2^\mu \,.
\eea
The polarized partonic matrix element squared involves 
the helicity combination $(+,+) - (+,-)$ with $+,-$ denoting the 
helicities $h_1$, $h_2$ of the incoming partons. 
Then from eq. (\ref{eq4}) we find 
\bea
\Delta~|M_{q \bar q \rightarrow Q\bar Q}|^2~=~\frac{ g^4}{4m^2}~ 
{\eta^\prime}^{\dagger} \sigma^i T^a {\xi^\prime}\, 
{\xi}^{\dagger} \sigma^j T^a {\eta}\,~[
2m^2 n_i n_j~ -~\delta_{ij} (k_1 \cdot k_2)]\,,
\label{eq9}
\eea
using $(k_2 \cdot L)_i=-(k_1 \cdot L)_i~=~m~n_i$. Here $n_i,n_j$ are the 
components of unit three-vectors ${\bf n}_1,{\bf n}_2$ 
which specify the polarizations of the heavy quarks and heavy
antiquarks respectively in the charmonium bound state. 
Their $z$-components are usually chosen along the beam direction. 
In a frame where $\bf{P}, \bf{k_1}$, and $\bf{k_2}$
are collinear, they are also collinear with the third components of 
${\bf n}_1,{\bf n}_2$. 
Taking the leading order term in an expansion in $q$ so that 
$P^2=4m^2$ we obtain
\bea
\Delta~|M_{q \bar q \rightarrow Q\bar Q}|^2~=~\frac{ g^4}{4}~ 
[ n_i n_j - \delta_{ij}]~ 
{\eta^\prime}^{\dagger} \sigma^i T^a {\xi^\prime} \, {\xi}^{\dagger} 
\sigma^j T^a {\eta}\,.
\label{eqqbar0}
\eea
Averaging over the initial color (by dividing by 9), we therefore get
\bea
\Delta~|M_{q \bar q \rightarrow Q\bar Q}|^2~=~\frac{ 4\pi^2 
\alpha_s^2}{9}~ 
[ n_i n_j - \delta_{ij}]~ 
{\eta^\prime}^{\dagger} \sigma^i T^a {\xi^\prime} 
{\xi}^{\dagger} \sigma^j T^a {\eta}\,.
\label{eqqbar1}
\eea
The two-component spinor factors can be identified with various heavy 
quarkonium bound states $H(\lambda)$ with different quantum numbers as 
follows, (see Appendix B in \cite{braaten})
\bea
&&~4m^2~ \eta^{\prime \dagger} \xi^\prime \,\xi^\dagger  \eta ~\equiv~
 < \chi^{\dagger} \psi \,P_{H(\lambda)}\,~\psi^\dagger  
 \chi >  ~=~\frac{4}{3} m <{\cal O}^H_1(^1S_0)>\,,
\nonumber \\
&&~4m^2~ \eta^{\prime \dagger} T^a \xi^\prime \,\xi^\dagger T^a \eta ~\equiv~
 < \chi^{\dagger} T^a \psi \,P_{H(\lambda)}\,~\psi^\dagger  
T^a \chi >  ~=~\frac{4}{3} m <{\cal O}^H_8(^1S_0)>\,,
\nonumber \\
&&~4m^2~ \eta^{\prime \dagger} \sigma^i \xi^\prime \,\xi^\dagger  \sigma^j 
\eta ~\equiv~
 < \chi^{\dagger} \sigma^i \psi \,P_{H(\lambda)}\,~\psi^\dagger  
\sigma^{j} \chi >  ~=~\frac{4}{3} U^\dagger_{\lambda i} U_{j\lambda } 
m <{\cal O}^H_1(^3S_1)>\,,
\nonumber \\
&&~4m^2~ \eta^{\prime \dagger} \sigma^i T^a \xi^\prime \,\xi^\dagger  
\sigma^j T^a \eta ~\equiv~
 < \chi^{\dagger} \sigma^i T^a \psi \,P_{H(\lambda)}\,~\psi^\dagger  
\sigma^{j} T^a \chi >  ~=~\frac{4}{3} U^\dagger_{\lambda i} U_{j\lambda } 
m <{\cal O}^H_8(^3S_1)>\,,
\nonumber \\
&&~4m^2~q^n {q^m}\eta^{\prime \dagger} \sigma^i \xi^\prime\, \xi^\dagger 
\sigma^j
\eta ~\equiv~ < \chi^{\dagger} (-\frac{i}{2}D^m) \sigma^i  \psi 
\,P_{H(\lambda)}\,~\psi^\dagger  
(-\frac{i}{2}D^n) \sigma^{j}  \chi >  \nonumber \\
&&~=~4 U^\dagger_{\lambda i}\,  U_{j\lambda } \delta^{mn} m 
<{\cal O}^H_1(^3P_0)>\,,
\nonumber \\
&&~4m^2~q^n {q^m}\eta^{\prime \dagger} \sigma^i T^a \xi^\prime 
\xi^\dagger  \sigma^{j} T^a \eta ~\equiv~ < \chi^{\dagger} 
(-\frac{i}{2}D^m) \sigma^i T^a \psi \,
P_{H(\lambda)}~\psi^\dagger  (-\frac{i}{2}D^n) \sigma^{j} T^a \chi >  
\nonumber \\
&&~=~4  U^\dagger_{\lambda i} \, U_{j\lambda } \delta^{mn} m 
<{\cal O}^H_8(^3P_0)>\,.
\label{had}
\eea
The helicity index $\lambda$ is a vector index in the spherical basis. 
The spherical basis states and cartesian basis states are related by a unitary 
transformation matrix $U_{\lambda i}$ which satisfies the relation
\bea
\sum_i U_{\lambda i} U^\dagger_{i \lambda}~=~1  
\nonumber \\
\sum_i U_{\lambda i} n^i=\delta_{\lambda 0}\,, 
\label{matr}
\eea
where $n^i$ is along the z-direction. Using the above relations 
we finally obtain
\bea
\Delta |M_{q \bar q \rightarrow H(\lambda)}|^2~ =~ -\frac{ 4\pi^2 
\alpha_s^2}{27}~ 
[1 - \delta_{\lambda 0}]~<{\cal{O}}^{H}_8(^3S_1)>\,.
\label{eq8}
\eea
The polarized quark-antiquark fusion process cross section is given by
\bea
\Delta \sigma_{q \bar q \rightarrow H(\lambda)}~ =~
-\delta(\hat s -4m^2)~\frac{ \pi^3 \alpha_s^2}{27m^3}~ 
[1-\delta_{\lambda 0}]~<{\cal{O}}^{H}_8(^3S_1)>\,,
\label{eq10}
\eea
and therefore vanishes for $\lambda=0$.

\subsection{The gg fusion process}

The matrix element for the gluon fusion process $g(k_1) + g(k_2) 
\rightarrow Q(p_1) + \bar{Q}(p_2)$ 
after including s, t, and u channel Feynman diagrams is given by
\bea
M_{gg  \rightarrow Q\bar Q}~=~-g^2~ \epsilon^a_\mu(k_1) 
\epsilon^{*b}_\nu(k_2) ~ 
[(\frac{1}{6}\delta^{ab} ~+~\frac{1}{2}d^{abc}T^c)~S^{\mu \nu}~+~\frac{i}{2}
f^{abc}T^c~F^{\mu \nu}]\,,
\label{eg1p}
\eea
where
\bea
S^{\mu \nu}~=~\bar{u}(p_1) 
[\frac{\gamma^\mu ~({{p}\!\!\!\slash}_1 - {{k}\!\!\!\slash}_1 +m)~\gamma^\nu}
{2p_1 \cdot k_1}~+~
\frac{\gamma^\nu ~({{p}\!\!\!\slash}_1 - {{k}\!\!\!\slash}_2 +m)~\gamma^\mu}
{2p_1 \cdot k_2}]\,
v(p_2)
\label{eg2}
\eea
and
\bea
&& F^{\mu \nu}~=~\bar{u}(p_1) 
[\frac{\gamma^\mu ~({{p}\!\!\!\slash}_1 - {{k}\!\!\!\slash}_1 +m)~\gamma^\nu}
{2p_1 \cdot k_1}~-~
\frac{\gamma^\nu ~({{p}\!\!\!\slash}_1 - {{k}\!\!\!\slash}_2 +m)~\gamma^\mu}
{2p_1 \cdot k_2} 
\nonumber \\
&& ~+~ \frac{2}{P^2}~V^{\mu \nu \lambda}(k_1,k_2,-k_1-k_2)~\gamma_\lambda ]
v(p_2)\,.
\label{eg3}
\eea
The three gluon vertex is denoted by $V^{\mu \nu \lambda}(k_1,k_2,k_3)~=~[
(k_1-k_2)^\lambda g^{\mu \nu}~+~(k_2-k_3)^\mu g^{\nu \lambda}~+~(k_3-k_1)^\nu 
g^{\lambda\mu} ]$.
Using various identities among the spinors and boost matrices 
from the appendix A of \cite{braaten} 
and after performing a considerable amount of algebra we find
\bea
&&\bar{u}(p_1) [\frac{\gamma^\mu ~{{k}\!\!\!\slash}_1~\gamma^\nu}
{2p_1 \cdot k_1}~+~
\frac{\gamma^\nu {{k}\!\!\!\slash}_2 ~\gamma^\mu}{2p_1 \cdot k_2}] v(p_2)~=~
\frac{i}{2m^2}(k_1-k_2)_\lambda \epsilon^{\rho \mu \nu \lambda} P_\rho 
\xi^\dagger \eta 
\nonumber \\
&& ~+~ \frac{(L\cdot k_1)_n}{m^3}~[P^\nu L^\mu_j-P^\mu L^\nu_j
+2g^{\mu \nu}(L\cdot k_1)_j-(k_1-k_2)^\mu
L^\nu_j-(k_1-k_2)^\nu L^\mu_j]~q^n \xi^\dagger \sigma^j \eta 
\nonumber \\
&& ~+~ \frac{(L\cdot k_1)_j}{m^3}~[P^\mu L^\nu_n-P^\nu L^\mu_n] ~q^n 
\xi^\dagger \sigma^j \eta 
~+~ \frac{1}{m}~[P^\mu L^\nu_j-P^\nu L^\mu_j] ~\xi^\dagger \sigma^j \eta\,, 
\label{eg4}
\eea
and
\bea
&&\bar{u}(p_1) [\frac{\gamma^\mu ~{{k}\!\!\!\slash}_1~\gamma^\nu}{2p_1 
\cdot k_1}~-~
\frac{\gamma^\nu {{k}\!\!\!\slash}_2 ~\gamma^\mu}{2p_1 \cdot k_2}] v(p_2)~=~
\frac{(L \cdot k_1)_n}{2m^4}(k_1-k_2)_\lambda 
\epsilon^{\rho \mu \nu \lambda} P_\rho q^n \xi^\dagger \eta 
\nonumber \\
&& ~-~ \frac{(L\cdot k_1)_n}{m^3}~[P^\nu L^\mu_j+P^\mu L^\nu_j]~q^n 
\xi^\dagger \sigma^j \eta
\nonumber\\
&& -\frac{1}{m}~[2g^{\mu \nu}(L\cdot k_1)_j-(k_1-k_2)^\mu 
L^\nu_j-(k_1-k_2)^\nu L^\mu_j]~\xi^\dagger \sigma^j \eta 
\nonumber \\
&&~+~ \frac{2}{m}~[L^\mu_n L^\nu_j-L^\nu_n L^\mu_j] ~q^n \xi^\dagger 
\sigma^j \eta\,. 
\label{eg5}
\eea
Hence 
\bea
&& S^{\mu \nu}~=~
\frac{i}{2m^2}(k_1-k_2)_\lambda \epsilon^{\rho \mu \nu \lambda} P_\rho 
\xi^\dagger \eta 
~+~ [\frac{(L\cdot k_1)_j}{m^3}~(P^\nu L^\mu_n-P^\mu L^\nu_n-2g^{\mu \nu}
(L\cdot k_1)_n) 
\nonumber \\
&& ~+~ \frac{2}{m}~[L^\mu_n L^\nu_j+L^\nu_n L^\mu_j]
+\frac{1}{m^3}(L \cdot k_1)_n[(k_1-k_2)^\mu L^\nu_j+(k_1-k_2)^\nu L^\mu_j]]~q^n 
\xi^\dagger \sigma^j \eta\,, 
\label{eg6}
\eea
which is symmetric under $k_1\leftrightarrow k_2$, $\mu \leftrightarrow \nu$
because $(k_1\cdot L)_i = - (k_2\cdot L)_i$,
and 
\bea
F^{\mu \nu}~=~
\frac{i(L \cdot k_1)_n}{2m^4}(k_1-k_2)_\lambda \epsilon^{\rho \mu \nu \lambda}
P_\rho q^n \xi^\dagger \eta 
+[k_2^\nu L^\mu_j-k_1^\mu L^\nu_j]~\xi^\dagger \sigma^j \eta \,, 
\label{eg7}
\eea
which is antisymmetric under the interchanges above.
For an incoming gluon with a helicity $\lambda_1$ the square of   
gluon polarization vector can be written as \cite{sivers}
\bea
\epsilon^a_\mu(k_1,\lambda_1) 
\epsilon^{*b}_\nu(k_1,\lambda_1)
~=~\frac{1}{2}~\delta^{ab}~
[-g_{\mu \nu}~+\frac{k_{1 \mu} k_{2\nu}+k_{2\mu}k_{1\nu}}{k_1 \cdot k_2}-
i\lambda_1 ~\epsilon_{\mu \nu \rho \delta} 
\frac{k_1^\rho k_2^\delta}{k_1 \cdot k_2}]\,,
\label{nopolsump}
\eea
with a similar result for the square of the second polarization vector 
$\epsilon^a_\mu(k_2,\lambda_2)\epsilon^{*b}_\nu(k_2,\lambda_2)$.
Using the relation
\bea
\epsilon_{\mu \mu^\prime \alpha \beta} ~k_1^\alpha k_2^\beta~=~2m^2 
\epsilon^{ijk} n_k L^\mu_i L^\nu_j\,,
\label{eps2}
\eea
from appendix A of \cite{braaten} and choosing longitudinally polarized gluons
we find that 
\bea
&&~ \Delta |M_{gg  \rightarrow Q\bar Q}|^2~=~-\frac{g^4}{4}~
\epsilon^{pqr} \epsilon^{p^\prime q^\prime r^\prime} n_r n_{r^\prime} 
\nonumber \\
&&~\times[S^{ab}S^{*ab} L_{\mu p} L_{\nu p^\prime} S^{\mu \nu} L_{\mu^\prime q} 
L_{\nu^\prime q^\prime} 
S^{* \mu^\prime \nu^\prime} +F^{ab}F^{*ab} L_{\mu p} L_{\nu p^\prime} 
F^{ \mu \nu}
L_{\mu^\prime q} L_{\nu^\prime q^\prime} F^{*\mu^\prime \nu^\prime}]\,,
\label{dm3}
\eea
where
\bea
S^{ab}~=~\frac{1}{6}\delta^{ab}~+~\frac{1}{2}d^{abc}T^c 
~~~~~~~{\rm and }~~~~~~~~~ F^{ab}~=~\frac{i}{2}f^{abc}T^c\,.   
\label{sf}
\eea
The cross terms between $S$ and $F$ vanish because
\bea
S^{ab}F^{*ab}~=~0~=~ S^{*ab}F^{ab}\,,
\eea
due to their symmetry properties.  Using various properties 
of the $L^\mu_i$ matrices from appendix A of \cite{braaten} 
and performing some lengthy algebra we finally 
obtain, after averaging over the initial color (by dividing by 64),
\bea
&& \Delta |M_{gg  \rightarrow Q\bar Q}|^2~=~-\frac{\pi^2 \alpha_s^2}{9}~[
\eta^{\prime \dagger} \xi^\prime \xi^\dagger \eta +
\frac{1}{m^2}[(n \cdot q) n_jq^\prime_{j^\prime}
 +(n \cdot q^\prime)n_{j^\prime} q_j -\frac{3}{2}(n \cdot q) 
(n \cdot q^\prime)n_j n_{j^\prime} 
\nonumber \\ 
&& -(n \times q^\prime)_j (n \times q)_{j^\prime}]
\eta^{\prime \dagger} \sigma^{j^\prime} \xi^\prime \xi^\dagger  
\sigma^{j} \eta +\frac{15}{8}
\eta^{\prime \dagger} T^a \xi^\prime \xi^\dagger T^a \eta +\frac{15}{8m^2} 
[(n \cdot q) n_jq^\prime_{j^\prime}
 +(n \cdot q^\prime)n_{j^\prime} q_j  
\nonumber \\
&& -\frac{3}{2}(n \cdot q) (n \cdot q^\prime)n_j n_{j^\prime} 
 -(n \times q^\prime)_j (n \times q)_{j^\prime}]
\eta^{\prime \dagger} \sigma^{j^\prime} T^a \xi^\prime \xi^\dagger  
\sigma^{j} T^a \eta 
\nonumber \\ 
&& +\frac{27}{8m^2}~
(n \cdot q) (n \cdot q^\prime)
\eta^{\prime \dagger} T^a \xi^\prime \xi^\dagger  T^a \eta ]\,. 
\label{dpol}
\eea
After identifying the different bound states as given in eq. (\ref{had}) 
and then using eq. (\ref{matr}) we obtain
\bea
&& \Delta |M_{gg  \rightarrow H(\lambda)}|^2~=~-\frac{\pi^2 
\alpha_s^2}{27}~[ <{\cal O}^H_1(^1S_0)> ~+~\frac{15}{8} 
<{\cal O}^H_8(^1S_0)> 
\nonumber \\
&& +\frac{3}{m^2} (\frac{1}{2} 
\delta_{\lambda 0}-1)[ <{\cal O}^H_1(^3P_0)>    
+\frac{15}{8} <{\cal O}^H_8(^3P_0)> ]~
+~\frac{81}{8m^2} <{\cal O}^H_8(^1P_1)> ]\,.
\label{dpolf}
\eea
Hence the partonic level inclusive cross sections for 
heavy quarkonium production with definite helicities 
$\lambda$ in polarized g-g collisions are given by
\bea
&& \Delta \sigma_{gg  \rightarrow H(\lambda)}~=~-\delta(\hat s -4m^2)
~\frac{\pi^3 \alpha_s^2}{108m^3}~[ <{\cal O}^H_1(^1S_0)> 
~+~\frac{15}{8} <{\cal O}^H_8(^1S_0)> 
\nonumber \\
&& +\frac{3}{m^2}
(\frac{1}{2} \delta_{\lambda 0}-1) 
~[ <{\cal O}^H_1(^3P_0)>~+~\frac{15}{8} <{\cal O}^H_8(^3P_0)> ]~
+~\frac{81}{8m^2} <{\cal O}^H_8(^1P_1)> ]\,.
\label{dplf}
\eea

\section{Results and Discussion}

Using the formulae derived above we compute the LO rapidity
distributions and spin asymmetries for the heavy charmonium systems
$J/\psi$ and $\psi^\prime$ in longitudinally polarized proton-proton 
collisions at RHIC. This provides interesting information on the 
polarization state of these heavy charmonium states.  In terms of the 
heavy quark relative velocity $v$, the non-perturbative matrix 
elements for ${\cal O}_8(^3P_J)$ and ${\cal O}_8(^3S_1)$
production scale like $v^7$ 
and for ${\cal O}_8(^1S_0)$ production scales like $v^6$ 
whereas those for ${\cal O}_8(^1P_1)$, ${\cal O}_1(^1S_0)$, 
production scale like $v^{10}$ 
and ${\cal O}_1(^3P_J)$ production scale like $v^{11}$ \cite{braaten,bodwin}.
Hence we will not include the latter contributions 
as they are expected to be small. In particular this means that 
contributions from processes in which the $Q \bar Q$ pair 
is produced in color-singlet states are not included.  
Folding eqs. (\ref{eq10}), (\ref{dplf}) with parton densities 
we find the following cross sections in 
longitudinally polarized proton-proton collisions
\bea
&& \Delta \sigma_{(pp \rightarrow J/\psi(\lambda) 
(\psi^\prime (\lambda)))}~=~\frac{\pi^3 \alpha_s^2}{27sm^3}~ 
\int_{4m^2/s}^1 \frac{dx_1 }{x_1}
~[\Delta f_q(x_1,2m) ~\Delta f_{\bar q}(\frac{4m^2}{x_1 s},2m)
\nonumber \\  
&& ~\times~ (\delta_{\lambda 0}-1)\,<{\cal{O}}^{J/\psi (\psi^\prime)}_8(^3S_1)> 
~+~\frac{15}{32}~\Delta f_g(x_1,2m) ~\Delta f_{g}(\frac{4m^2}{x_1 s},2m)
\nonumber \\
&&~\times~[ \frac{9}{m^2} (1-\frac{1}{2} \delta_{\lambda 0}) 
<{\cal O}^{J/\psi (\psi^\prime)}_8(^3P_0)> 
~-~<{\cal O}^{J/\psi (\psi^\prime) }_8(^1S_0)> ] ]\,,
\label{polfin}
\eea
where $\Delta f (x,Q)(\Delta g(x,Q))$ denote the polarized quark (gluon) 
distribution functions inside the proton at the scale $Q$. 
The corresponding production cross sections
for unpolarized proton-proton collisions are \cite{braaten}:
\bea
&& \sigma_{(pp \rightarrow J/\psi(\lambda) 
(\psi^\prime (\lambda)))}~=~\frac{\pi^3 \alpha_s^2}{27sm^3}~ 
\int_{4m^2/s}^1 \frac{dx_1 }{x_1}
~[f_q(x_1,2m) ~f_{\bar q}(\frac{4m^2}{x_1 s},2m)
\nonumber \\ 
&&~\times~(1-\delta_{\lambda 0})~<{\cal{O}}^{J/\psi (\psi^\prime)}_8(^3S_1)> 
~+~\frac{15}{32}~f_g(x_1,2m) ~ f_{g}(\frac{4m^2}{x_1 s},2m)
\nonumber \\
&&~\times~[ \frac{9}{m^2} (1-\frac{2}{3} \delta_{\lambda 0}) 
<{\cal O}^{J/\psi (\psi^\prime)}_8(^3P_0)> ~+~
<{\cal O}^{J/\psi (\psi^\prime) }_8(^1S_0)> ]]\,.
\label{unpolfin}
\eea
The spin asymmetry $A_{LL}(\lambda)$ is given by the ratio of the 
above cross sections
\bea
A_{LL}(\lambda)~=~\frac{d \Delta \sigma(\lambda)}{d \sigma(\lambda)}\,.
\label{spinasym}
\eea

\begin{figure}[htb]
\vspace{2pt}
\centering{\rotatebox{270}{\epsfig{figure=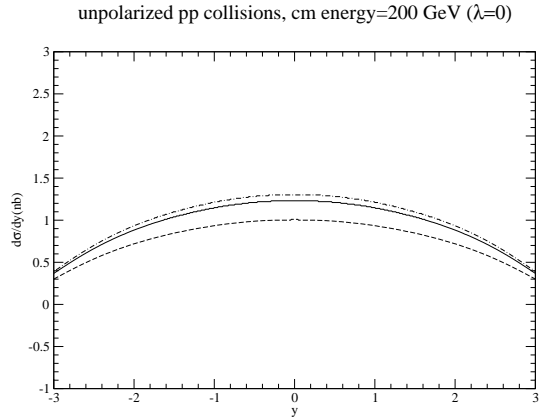,height=7cm}}}
\caption{Differential rapidity distributions for $J/\psi$ production
at $\sqrt s$ = 200 GeV.  The solid, dashed and dot-dashed lines, each to 
be multiplied by a scale factor of 200, are for unpolarized p-p collisions
with $J/\psi$ helicities $\lambda$ = 0 in scenarios 1, 2 and 3 respectively.}
\label{fig1}
\end{figure}

We begin with a discussion of the rapidity distribution for $J/\psi$
production. Color octet contributions have been obtained from an analysis of 
charmonium transverse momenta differential cross section data
from the Fermilab Tevatron, see \cite{BK} and \cite{CL}.
In particular central values are known for  
$<{\cal{O}}^{J/\psi}_8(^3S_1)>$ and the combination
\bea
M^{J/\psi}(^1S_0^{(8)},{}^3P_0^{(8)})=
<{\cal{O}}^{J/\psi}_8(^1S_0)> + 3.5 <{\cal{O}}^{J/\psi}_8(^3P_0)>/m^2\,,
\eea
together with reasonable error ranges from both statistical and theoretical
uncertainties. Measurements of the polarized
cross sections should allow one to determine the individual values for
these contributions. All we can do at present is assume plausible 
values to guide experimental investigations.   

\begin{figure}[htb]
\vspace{2pt}
\centering{\rotatebox{270}{\epsfig{figure=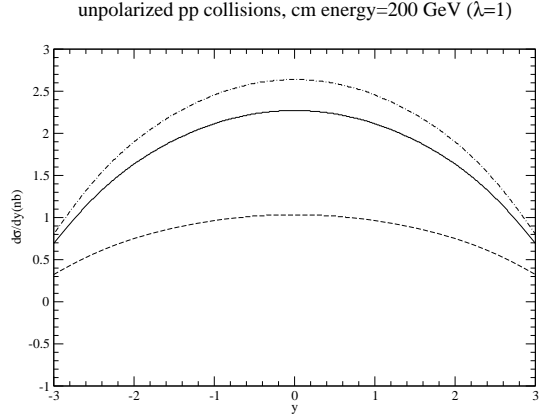,height=7cm}}}
\caption{Differential rapidity distributions for $J/\psi$ production
at $\sqrt s$ = 200 GeV.  The solid, dashed and dot-dashed lines, each to 
be multiplied by a scale factor of 200, are for unpolarized p-p collisions
with $J/\psi$ helicities $\lambda$ = 1 in scenarios 1, 2 and 3 respectively.}
\label{fig2}
\end{figure}

Therefore we choose three scenarios.\\
1.\qquad  
$<{\cal{O}}^{J/\psi}_8(^1S_0)>$ and $<{\cal{O}}^{J/\psi}_8(^3P_0)>/m^2$
are roughly equal, so we set both to 0.0087 GeV${}^3$.\\
2. \quad
$<{\cal{O}}^{J/\psi}_8(^1S_0)>$ is much larger than
$<{\cal{O}}^{J/\psi}_8(^3P_0)>/m^2$
so we set the former to 0.039 GeV${}^3$ and the latter to zero.\\
3.\qquad
$<{\cal{O}}^{J/\psi}_8(^1S_0)>$ is much smaller than
$<{\cal{O}}^{J/\psi}_8(^3P_0)>/m^2$
so we set the former to zero and the latter to 0.01125 GeV${}^3$.\\
In all three cases 
$M^{J/\psi}(^1S_0^{(8)},{}^3P_0^{(8)})\approx 0.039\, {\rm GeV}^3$,
which is the central value in \cite{BK} for GRV leading order (LO)
parton densities \cite{grv95}.  We also take 
$ <{\cal{O}}^{J/\psi}_8(^3S_1)> = 0.0112\, {\rm GeV}^3$ in agreement
with the value in \cite{BK}. Note that these values can still vary up
and down by approximately fifty percent.  

\begin{figure}[htb]
\vspace{2pt}
\centering{\rotatebox{270}{\epsfig{figure=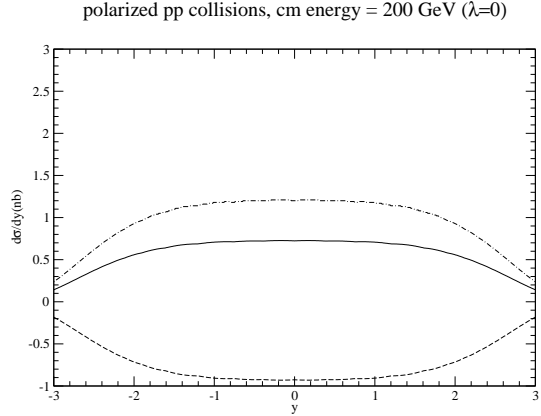,height=7cm}}}
\caption{Differential rapidity distributions for $J/\psi$ production
at $\sqrt s$ = 200 GeV.  The solid, dashed and dot-dashed lines
are for polarized p-p collisions with $J/\psi$ helicities $\lambda$ = 0 
in scenarios 1, 2 and 3 respectively.}
\label{fig3}
\end{figure}

The central arm (forward arm) electron (muon) detector at the PHENIX 
experiment covers the $J/\psi$ rapidity range $-0.5<~y~<0.5$ ($1<~|y|~<2$).
We will present our differential rapidity distributions and spin asymmetries 
for $J/\psi$ and $\psi^\prime$ production with 
helicities $\lambda$ = 1 and 0 in unpolarized and polarized p-p
collisions at $\sqrt s$ = 200 GeV and 500 GeV in the above detector 
acceptance ranges. 

\begin{figure}[htb]
\vspace{2pt}
\centering{\rotatebox{270}{\epsfig{figure=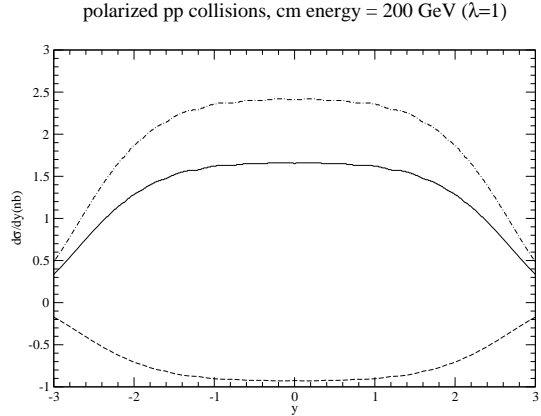,height=7cm}}}
\caption{Differential rapidity distributions for $J/\psi$ production
at $\sqrt s$ = 200 GeV.  The solid, dashed and dot-dashed lines
are for polarized p-p collisions with $J/\psi$ helicities $\lambda$ = 1 
in scenarios 1, 2 and 3 respectively.}
\label{fig4}
\end{figure}

We take the charm quark mass $m$=1.5 GeV and the mass factorization scale
equal to $2 m$. Several groups have produced polarized parton density sets
\cite{GS},\cite{grsv} and \cite{blbo}.  We choose the GRV unpolarized LO 
parton densities \cite{grv98} and the GRSV \cite{grsv} polarized densities.
The latter authors have a standard scenario and a valence scenario. 
For simplicity we choose the former. Therefore we always use the LO four 
flavour sets (for the u, d, s and g partons)
and we set $n_f=4$ in the one-loop running coupling constant and the 
parton densities.  For both parton density sets  we use
$\Lambda^{\rm LO}_4 = 175 $ MeV, so that 
$\alpha_s^{\rm LO}(m_Z) = 0.121$ at the mass of the Z. 

\begin{figure}[htb]
\vspace{2pt}
\centering{\rotatebox{270}{\epsfig{figure=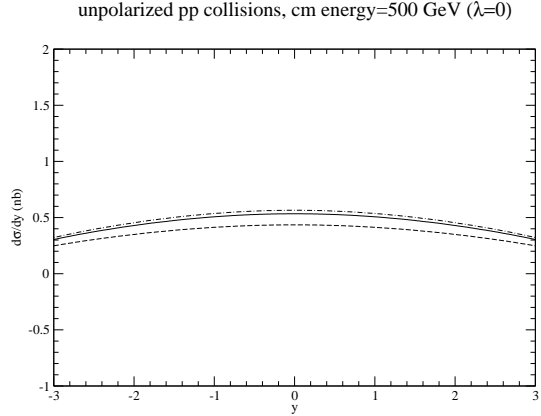,height=7cm}}}
\caption{Differential rapidity distributions for $J/\psi$ production
at $\sqrt s$ = 500 GeV.  The solid, dashed and dot-dashed lines, each to 
be multiplied by a scale factor of 1000, are for unpolarized p-p collisions
with $J/\psi$ helicities $\lambda$ = 0 in scenarios 1, 2 and 3 respectively.}
\label{fig5}
\end{figure}

We first choose helicity $\lambda=0$ which eliminates contributions from
$<{\cal{O}}^{J/\psi}_8(^3S_1)>$ in eqns. (\ref{polfin}) 
and (\ref{unpolfin}).  Note that this contribution only appears in the
quark-antiquark channel.   
In Fig. 1 we present the rapidity differential distributions 
for $J/\psi$ production with $\lambda=0$ in unpolarized p-p collisions 
at $\sqrt s$ = 200 GeV. The solid, dashed and dot-dashed lines, each to 
be multiplied by 200, are the differential distributions for the scenarios 
1, 2 and 3 respectively. As expected there is very little difference 
between them. Note that the $y$ axis is taken from -1 to +3 as in the
case of the polarized plots, which will be shown shortly.
Fig. 2 contains the corresponding distributions when
$\lambda=1$. Now $<{\cal{O}}^{J/\psi}_8(^3S_1)>$ contributes to 
eqn. (\ref{unpolfin}) and is responsible for the tiny difference
between the results for the second scenarios (the dashed lines). 
From this we conclude that the gluon-gluon contributions completely 
dominate the quark-antiquark contribution for this cm energy. 
The larger differences between the other results are 
due to the changes in the prefactors multiplying the octet contributions
in eqn. (\ref{unpolfin}).

\begin{figure}[htb]
\vspace{2pt}
\centering{\rotatebox{270}{\epsfig{figure=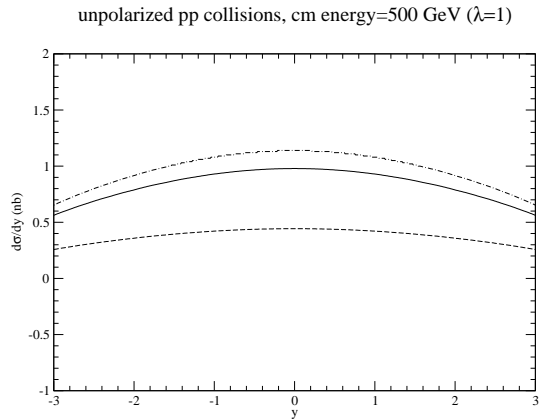,height=7cm}}}
\caption{Differential rapidity distributions for $J/\psi$ production
at $\sqrt s$ = 500 GeV.  The solid, dashed and dot-dashed lines, each to 
be multiplied by a scale factor of 1000, are for unpolarized p-p collisions
with $J/\psi$ helicities $\lambda$ = 1 in scenarios 1, 2 and 3 respectively.}
\label{fig6}
\end{figure}

We now give the corresponding results for longitudinally polarized
p-p collisions at the same cm energy using eqn. (\ref{polfin}). 
Fig. 3 has $\lambda=0$ and Fig. 4 has $\lambda=1$. 
In both cases scenario 2, where $<{\cal{O}}^{J/\psi}_8(^3P_0)>/m^2 =\, 0$, 
yields, as expected from the signs in eqn. (\ref{polfin}), negative results.
We conclude that the polarized results are at least a factor of
200 lower then the unpolarized ones but they could be much smaller
due to cancellations between the color octet contributions
in eqn. (\ref{polfin}) and could possibly be negative.

\begin{figure}[htb]
\vspace{2pt}
\centering{\rotatebox{270}{\epsfig{figure=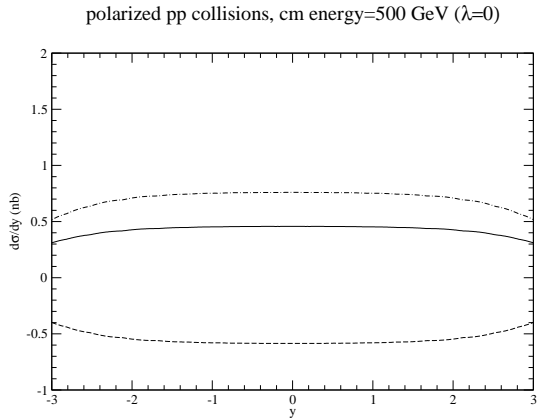,height=7cm}}}
\caption{Differential rapidity distributions for $J/\psi$ production
at $\sqrt s$ = 500 GeV.  The solid, dashed and dot-dashed lines
are for polarized p-p collisions with $J/\psi$ helicities $\lambda$ = 0 
in scenarios 1, 2 and 3 respectively.}
\label{fig7}
\end{figure}

In Figs. 5 - 8 we repeat the plots in Figs. 1 - 4 but for 
$\sqrt{s} = 500$ GeV. This time we divide the unpolarized values
by 1000 so that they fit on the same scales as the polarized ones.
Also we choose the $y$-axis from -1 to +2 so that the polarized and
unpolarized plots can be easily compared. The comments we made
earlier about Figs. 1 - 4 are also valid for these plots.
Here we see that that the polarized values are at least a factor of
1000 times smaller than the unpolarized ones, which is due to the rapid rise
of the unpolarized gluon density at small $x$,
but again they could be even smaller due 
to cancellations between the color octet contributions.

\begin{figure}[htb]
\vspace{2pt}
\centering{\rotatebox{270}{\epsfig{figure=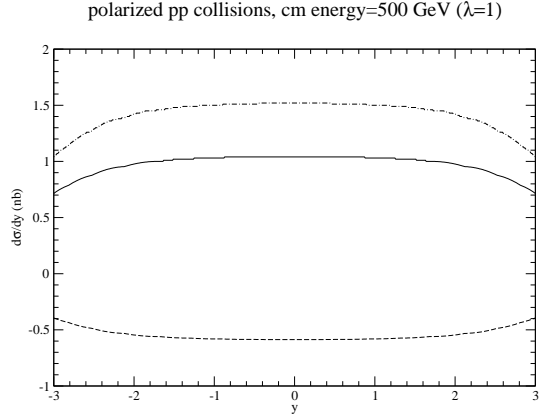,height=7cm}}}
\caption{Differential rapidity distributions for $J/\psi$ production
at $\sqrt s$ = 500 GeV.  The solid, dashed and dot-dashed lines
are for polarized p-p collisions with $J/\psi$ helicities $\lambda$ = 1 
in scenarios 1, 2 and 3 respectively.}
\label{fig8}
\end{figure}

For completeness we give in Fig.9 the rapidity distributions for the spin 
asymmetry $A_{LL}$ defined in eq. (\ref{spinasym}) for $J/\psi$ production 
in p-p collisions at $\sqrt s$ = 200 GeV. This is for the scenario 1.
The solid and dotted lines are for helicities 
$\lambda=1$ and 0 respectively. We see that the ratios are rather flat over
the central rapidity range. For scenario 2 $A_{LL}$ is negative and roughly
the same in absolute value. For scenario 3 $A_{LL}$ is positive and
about the same. Therefore we do not show the latter two cases.   

\begin{figure}[htb]
\vspace{2pt}
\centering{\rotatebox{270}{\epsfig{figure=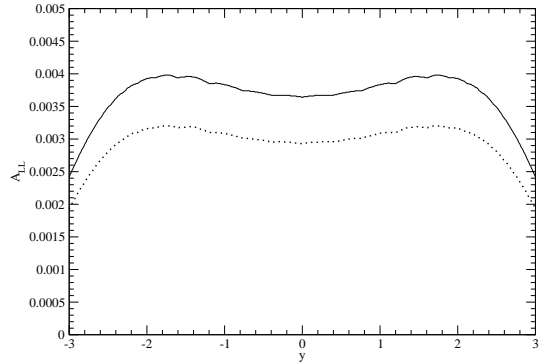,height=7cm}}}
\caption{Differential rapidity distributions for $A_{LL}$ 
at $\sqrt s$ = 200 GeV in scenario 1.  The solid and dotted lines
are for helicities $\lambda=1$ and 0 respectively.}
\label{fig9}
\end{figure}

Finally Fig. 10 contains the rapidity distributions of the spin 
asymmetries $A_{LL}$ for $J/\psi$ production in  
p-p collisions at $\sqrt s$ = 500 GeV.  The solid and dotted lines are 
the spin asymmetries for helicities $\lambda=1$ and 0 respectively.
The ratio is also rather flat and smaller than in Fig. 9, which is due
to the increase in the unpolarized gluon density at small $x$.
We do not give the plots for scenarios 2 and 3 as they are similar 
in absolute magnitude. 

Let us now consider $\psi^\prime$ production. 
Central values for  
$ <{\cal{O}}^{\psi^\prime}_8(^3S_1)>$ and the combination
\bea
M^{\psi^\prime}(^1S_0^{(8)},{}^3P_0^{(8)})=
<{\cal{O}}^{\psi^\prime}_8(^1S_0)> 
+ 3.5 <{\cal{O}}^{\psi^\prime}_8(^3P_0)>/m^2\,,
\eea
together with reasonable error ranges from both statistical and theoretical
uncertainties are given in \cite{BK}. 
However we note that only the former, equal to 0.0046 GeV${}^3$, 
is derived from data analysis while the latter is not. The data were simply 
not good enough so it was assumed that the ratios of the above
color combinations are the same for the $J/\psi$ and $\psi^\prime$
respectively. However we have seen that the contribution from
$ <{\cal{O}}^{\psi^\prime}_8(^3S_1)>$ in the quark-antiquark 
channel is completely negligible compared to the other color octet 
contributions from the gluon-gluon channel in this energy range. 
Hence we only need to look at the ratios 
in \cite{BK} to determine the reduction factor for the
rapidity distributions for $\psi^\prime$ production. This is 3.5 
so all plots given above in Figs. 1 - 8 can be used 
for $\psi^\prime$ production by simply dividing them by this number. The 
$A_{LL}$ ratio plots in Figs. 9 and 10 are not affected since this 
factor cancels between the numerator and denominator.

In this paper we have calculated inclusive production cross sections and 
spin asymmetries for heavy quarkonium states
with definite helicities in polarized proton-proton collisions using the 
non-relativistic QCD color-octet mechanism. 
We have presented the LO results for $J/\psi $ and $\psi^\prime$
rapidity differential distributions with definite
helicities in polarized p-p collisions at $\sqrt s$ = 200 GeV and 500 GeV 
at the RHIC within the PHENIX detector acceptance range. One can see from the 
figures that the contributions from the $\lambda = \pm 1$ states dominate.
This is explained by the fact that the RHIC is a p-p collider so the 
contribution from the quark-antiquark channel is small compared to the
contribution from the gluon-gluon channel and the coefficient in front
of the $<{\cal{O}}^{\psi^\prime}_8(^3P_0)>$ term in eqn. (\ref{polfin})
is larger for $\lambda=1$ than for $\lambda=0$. 
The PHENIX experiment should be able to measure these spin 
asymmetries.  The study of heavy quarkonium production
with definite helicities in polarized p-p collisions is unique because 
it tests the spin transfer processes in perturbative QCD.  As Tevatron 
data for heavy quarkonium polarization \cite{expt} is not explained by 
the color octet mechanism \cite{CDFpolarization}, it will be interesting 
to compare these theoretical results with the future experimental
data at RHIC to test the NRQCD color octet heavy quarkonium production 
mechanism with respect to polarization. 
A calculation of the $p_T$ distribution in heavy quarkonium
production with definite polarization states in polarized p-p 
collisions at RHIC should also show interesting features.

\begin{figure}[htb]
\vspace{2pt}
\centering{\rotatebox{270}{\epsfig{figure=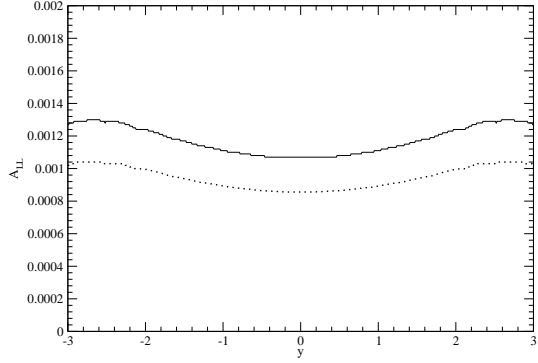,height=7cm}}}
\caption{Differential rapidity distributions for $A_{LL}$ 
at $\sqrt s$ = 500 GeV in scenario 1.  The solid and dotted lines
are for helicities $\lambda=1$ and 0 respectively.}
\label{fig10}
\end{figure}

\acknowledgments
We thank Ming X. Liu and George Sterman for discussions.  This work was 
supported in part by the National Science Foundation, grants PHY-0071027, 
PHY-0098527, PHY-0354776 and PHY-0345822.


\begin{thebibliography}{aaaaaaa}
\bibitem{qgp} Proceedings of the Quark Matter conference,
August 4-9, (2005), Budapest, Hungary;\\ 
M. Gyulassy and L. McLerran, Nucl. Phys. A750, 30 (2005), nucl-th/0405013;\\ 
G. C. Nayak, A. Dumitru, L. McLerran and W. Greiner, Nucl. Phys. 
A687, 457 (2001), hep-ph/0001202;\\ 
F. Cooper, E. Mottola and G. C. Nayak, Phys. Lett. B555, 181 (2003),
hep-ph/0210391;\\
R. S. Bhalerao and G. C. Nayak, Phys. Rev. C61, 054907 (2000),
hep-ph/9907322;\\ 
G. G. Nayak and V. Ravishankar, Phys. Rev. C58 (1998) 356; Phys. Rev. D55,
6877 (1997), hep-ph/9610215.
\bibitem{spin} V. Ravindran, J. Smith and W. L. van Neerven, 
Nucl. Phys. B682, 421 (2004), hep-ph/0311314;
Nucl. Phys. B647, 275 (2002), hep-ph/0207076 ; 
Nucl. Phys. Proc. Suppl. 135, 14 (2004), hep-ph/0405233;\\ 
W. Vogelsang and F. Yuan, hep-ph/0507266;\\ 
W. Vogelsang, Pramana 63, 1251 (2004), hep-ph/0405069. 
\bibitem{phenix}
H. P. Da Costa (for the PHENIX collaboration)
"Phenix results for $J/\psi$ production in Au+Au and Cu+Cu collisions
at $\sqrt{S_{\rm NN}}= 200$ GeV, proceedings of the Quark Matter conference,
August 4-9, (2005), Budapest, Hungary, http://qm2005.kfki.hu/; 
I. Younus, Hawaii DNP2005 APS/JPS meeting. 
\bibitem{QWGBrambilla} 
N.\ Brambilla {\it et al.} (Quarkonium
Working Group), hep-ph/0412158, and references therein.
\bibitem{bodwin} 
G. T. Bodwin, E. Braaten and G. P. Lepage, Phys. Rev. D51, 1125 (1995),
Erratum-ibid D55, 5853 (1997),
hep-ph/9407339.
\bibitem{nayak} 
G. C. Nayak, J-W. Qiu and G. Sterman, Phys. Lett. B613, 45 (2005), 
hep-ph/0501235;\\
G. C. Nayak, J-W. Qiu and G. Sterman, Stony Brook Preprint, 
hep-ph/0509021.
\bibitem{CDFoctet}
E. Braaten and S. Fleming, Phys.\ Rev.\ Lett.\ 74, 3327 (1995),
hep-ph/9411365;\\ 
E. Braaten, S. Fleming and T. C. Yuan, 
Ann.\ Rev.\ Nucl.\ Part.\ Sci.\ 46, 197 (1996), hep-ph/9602374;\\ 
E. Braaten, S. Fleming and A. K. Leibovich, Phys.\ Rev.\ D63, 094006 (2001),
hep-ph/0008091.
\bibitem{CL}
P. L. Cho and A. K. Leibovich, Phys.\ Rev.\ D53, 6203 (1996), hep-ph/9511315; 
Phys.\ Rev.\ D53, 150 (1996), hep-ph/9505329. 
\bibitem{HERAoctet} 
M. Cacciari and M. Kramer, Phys. Rev. Lett. 76, 4128 (1996),
hep-ph/9601276;\\
M. Beneke, M. Kramer and M. Vanttinen, Phys. Rev. D57, 4258 (1998),
hep-ph/9709376;\\
J. Amundson, S. Fleming and I. Maksymyk, Phys. Rev. D56, 5844 (1997),
hep-ph/9601298;\\
R. M. Goodbole, D. P. Roy and K. Sridhar, Phys. Lett. B373, 328 (1996),
hep-ph/9511433;\\
B. A. Kniehl and G. Kramer, Phys. Rev. D56, 5820 (1997), 
hep-ph/9706369.
\bibitem{LEPoctet} 
C. G. Boyd, A. K. Leibovich and I. Z. Rothstein,
Phys. Rev. D59, 054016 (1999), hep-ph/9810364;\\
M. Klasen, B. A. Kniehl, L. N. Mihaila
and M. Steinhauser, Phys. Rev. Lett. 89, 032001 (2002), hep-ph/0112259.
\bibitem{fixedtargetoctet} 
M. Beneke and I. Z. Rothstein, Phys. Rev. D54, 2005 (1996) 
[Erratum-ibid. D54, 7082] (1996)], hep-ph/9603400;\\
W. K. Tang and M. Vanttinen, Phys. Rev. D54, 4349 (1996), hep-ph/9603266;\\
S. Gupta and K. Sridhar, Phys. Rev. D54, 5545 (1996), hep-ph/9601349.
\bibitem{cooper} F. Cooper, M. X. Liu and G. C. Nayak, 
Phys. Rev. Lett. 93, 171801 (2004), hep-ph/0402219; \\
G. C. Nayak, M. X. Liu and F. Cooper, Phys. Rev. D68, 034003 (2003),
hep-ph/0302029.
\bibitem{pol} 
M. Klasen, B. A. Kniehl, L. N. Mihaila and M. Steinhauser,
Phys. Rev. D68, 034017 (2003), hep-ph/0306080.
\bibitem{fm} S. Fleming and I. Maksymyk, Phys. Rev. 54 (1996) 3608, 
hep-ph/9512320.
\bibitem{gm}
S. Gupta and P. Mathews, Phys. Rev. D55, 7144 (1997), hep-ph/9609504;
Phys. Rev. D56, 3019 (1997), hep-ph/9703370; 
Phys. Rev. D56, 7341 (1997), hep-ph/9706541.
\bibitem{braaten} E. Braaten and Y-Q Chen, Phys. Rev. D54, 3216 (1996),
hep-ph/9604237.
\bibitem{expt} T. Affolder, et al, CDF Collaboration,
Phys. Rev. Lett. 85, 2886 (2000), 
hep-ex/0004027.
\bibitem{CDFpolarization} 
E. Braaten, B. A. Kniehl and J. Lee, Phys. Rev. D62, 094005 (2000),
hep-ph/9911436;\\ 
E. Braaten and J. Lee, Phys. Rev. D63, 071501 (2001),
hep-ph/0012244;\\
M. Beneke and M. Kraemer, Phys. Rev. D55, 5269 (1997),
hep-ph/9611218;\\
A. K. Leibovich, Phys. Rev. D56, 4412 (1997),
hep-ph/9610381.
\bibitem{sivers} J. Babcock, E. Monsay, and D. Sivers, 
Phys. Rev. D19, 1483 (1979);\\
V. Ravindran, J. Smith and W. L. van Neerven, 
Nucl. Phys. B682, 421 (2004), hep-ph/0311304. 
\bibitem{BK}
M. Beneke and M. Kr\"amer, Phys. Rev. D55, 5269 (1997),
hep-ph/9611218.
\bibitem{grv95}
M. Gl\"uck, E. Reya and A. Vogt, Z. Phys. C67, 433 (1995).
\bibitem{GS}
T. Gehrmann and W.J. Stirling, Phys. Rev. D53, 6100 (1996),
hep-ph/9512406.  
\bibitem{grsv}
M. Gl\"uck, E. Reya, M. Stratmann and W. Vogelsang, Phys. Rev. D63, 094005
(2001), hep-ph/0011215.
\bibitem{blbo}
J. Blumlein and H. B\"ottcher, Nucl. Phys. B636, 225 (2002),
hep-ph/0203155.
\bibitem{grv98}
M. Gl\"uck, E. Reya and A. Vogt, Euro. Phys. J. C5, 461 (1998),
hep-ph/9806404.
\end{thebibliography}
\end{document}